\documentclass[a4paper]{jpconf}
\usepackage{graphicx}
\usepackage[numbers,square,sort&compress]{natbib}
\usepackage{xcolor}
\usepackage[section]{placeins}
\usepackage{booktabs, array}
\usepackage{multirow} 
\usepackage{listings}
\usepackage{hyperref}
\def \cobj #1{\lstinline[basicstyle=\ttfamily]{#1}}

\usepackage{lineno}

\selectcolormodel{rgb}
\definecolor{LHCb dark}{rgb}{0.0000,0.3412,0.6549}
\definecolor{UC red}{rgb}{0.8196,0.1176,0.2314} 
\definecolor{brickred}{rgb}{0.8, 0.25, 0.33}

\begin{document}

\title{MCBooster: a library for fast Monte Carlo generation of 
phase-space decays {on}  massively parallel platforms.}

\author{A A Alves J\'unior$^1$ and M D Sokoloff$^2$ }

\address{Physics Department ML0011l University of Cincinnati;  Cincinnati, OH 45221}

\ead{aalvesju@cern.ch$^1$ and sokoloff@ucmail.uc.edu$^2$ }

\begin{abstract}

MCBooster is a header-only, C++11-compliant library that provides 
routines to generate and perform calculations on large samples of 
phase space Monte Carlo events.  
To achieve superior performance, MCBooster is capable to perform most of its 
calculations in parallel using CUDA-
and OpenMP-enabled devices. 
MCBooster is {built} on top of the Thrust library and runs on Linux systems. 
This contribution summarizes the main features of MCBooster. 
A basic description of the user interface and some examples of applications are 
provided, along with measurements of performance in a variety of environments

\end{abstract}

\section{Introduction}

Phase space Monte Carlo simulates the 
decay kinematics of a mother particle decaying into $ N $ daughter particles 
with no intermediate resonances. 
Samples of phase space Monte Carlo events are widely used in 
high energy physics for calculating phase volumes.
They are also used  as the starting point to implement and  describe decays 
containing one or more intermediate resonances and
to simulate the response of detectors \cite{James:275743}. 

MCBooster is a header-only, C++11-compliant library for the generation of large 
samples of phase-space Monte Carlo events on massively parallel platforms. 
It was released on GitHub \footnote{https://github.com/MultithreadCorner/MCBooster} in the spring of 2016. 
The core libraries implement the Raubold-Lynch algorithm \cite{James:275743};
they are able to generate the full kinematics of decays with up to nine particles in the final state.
{MCBooster} supports the generation of sequential decays as well as the 
parallel evaluation of arbitrary functions over the generated events.
The output of MCBooster  accords with popular and well-tested 
software packages such as GENBOD (W515 from CERNLIB) \cite{CERNLIB} and TGenPhaseSpace 
in the ROOT framework \cite{Brun:1997pa}. 
MCBooster is built on top of the Thrust library and runs on Linux systems. 
It deploys transparently on NVidia CUDA-enabled GPUs as well as
multicore CPUs.

This contribution summarizes the main features of MCBooster. A basic 
description of the user interface is presented in \autoref{sec:implementation}.
Some examples of applications are provided in \autoref{sec:worked_example}.
Measurements of performance on a variety of
platforms and in a variety of environments are discussed in \autoref{sec:perf}.

\section{Phase space Monte Carlo}
\label{sec:php}
Considering {conservation of energy and momentum,}
and the relativistic mass-energy relation {($ m^2 = E^2 - p^2 $)}, 
the kinematics of the decay of a mother particle {at rest to}
an {N}-particle final state
can be described by $4N - 4 - 3 - N = 3N -7 $ independent parameters
{($ 4N $ for the four-momenta of the $ N $ daughters, 
4 for conservation of energy and momentum, 3 for angles in the
mother's center-of-momentum about which the daughters' momenta can be rotated
without changing the kinematics, and $ N $ for the  mass-energy relations
of the daughters).}

The Raubold-Lynch {algorithm} is  described in detail on \cite{James:275743}. 
Basically, the method {models} 
{an N}-body decay as a recursive series of two-body 
decays \cite{James:275743, Byckling_Kajantie_1973}.
The four-momenta of the final state particles are randomly generated and are kinematically consistent. 
All events are accepted, and a weight proportional to the local 
phase space density is calculated in order to ensure the correct event distribution. 
The resulting sample can be unweighted in parallel using one of the methods provided by the package. 

\section{Implementation highlights}
\label{sec:implementation}

MCBooster is implemented on top of the Thrust library as a header-only, C++11-compliant library
and runs on Linux systems. 
{It} supports systems
compatible with NVidia CUDA and OpenMP. 
{In each call to the principal method, events}
are generated in large bunches (typically
multiples of a million), depending on the memory
available on the device.

Some of the main features of MCBooster are

\begin{itemize}
\item The complete set of four-vectors is available for the
final states. MCBooster can output weighted and
{unweighted} samples.
\item Interfaces exist for intrusive and non-intrusive evaluation
of arbitrary functions of the phase space coordinates.
\item The MCBooster interface also supports the generation of
sequential decays.
\end{itemize}

Three- and four-vectors are described by the classes \cobj{Vector3R} and \cobj{Vector4R}.
Event generation is managed by the class  \cobj{PhaseSpace}, which is configured via the constructor
with the masses of the particles and the number of events to generate. 
The methods \cobj{PhaseSpace::Generate(Vector4R mother)} and  \cobj{PhaseSpace::Generate(MCBooster::Particles mothers)}
perform the actual event generation. Generated events are kept on the device. 
The containers have the lifetime of the corresponding \cobj{MCBooster::PhaseSpace} object. 
Allocated memory can be reused in subsequent runs. 
The user can export the {generated events} from {the}
device to {the} host and store {them} 
in the dedicated container class 
\cobj{Events}.
  
MCBooster provides three interfaces to perform the evaluation of arbitrary functions 
over the generated events in parallel. 
To use such interfaces, it is {necessary only}
to derive a functor from the corresponding interface and 
to implement the corresponding operator().

\section{A worked example} 
\label{sec:worked}

To exemplify some of the basic functionality of MCBooster, 
the code snippets below show
how to generate 10 million events and calculate some parameters corresponding 
to the decay chain $B^0\to K^{\ast} J/\psi$, 
 with $K^{\ast}\to K^+ \pi^-$ {and}  $J/\psi\to\mu^+\mu^-$. 
The parameters are

  \begin{itemize}
   \item $M(K,\pi)$, the invariant mass of the $K\pi$ system; 
   \item $\cos(\theta_{K})$, the {helicity}  angle of the $K^{\ast}$;
   \item $\cos(\theta_{\mu})$, the {helicity} angle of the $J/\psi$;
   \item $\Delta\phi$ difference between the decay planes of the $K^{\ast}$ and $J/\psi$.
  \end{itemize}
 
Given it's very narrow width, the invariant mass of the {$ J/\psi $} is fixed at its nominal value, 
{an excellent approximation given its} very narrow {width},
one can use {the} two-body  {$ K \pi $}
invariant mass plus the angular variables to {fully} specify 
the kinematic state of each generated event.

\subsection{Generation of the events}

\begin{center}
 \begin{lstlisting}[caption={C++ code listing showing how to configurate, generate and export to the host memory 10 million events of $B^0 \to K^+\pi^- J\psi$.} ,
 label=lst:BKPiJpsi_snipet, float, basicstyle=\tiny,linewidth=.90\textwidth]
	// Some input 
	GLong_t nEvents = 10000000; //10 Million events
	GReal_t motherMass = 5.2795;
                
	// Masses of B0 daugthers {J/psi, kaon , pion}
	vector<GReal_t> daughterMasses
         ={3.096916,0.493677, 0.13957018};
	
	// Constructor
	PhaseSpace phspB0( motherMass, daughterMasses, nEvents);
	
	// Actual generation (all decays from same mother particle)
	phspB0.Generate( Vector4R( motherMass, 0.0, 0.0, 0.0) );
	
	// Event container 
	Events *MyEvents = new Events(daughterMasses.size(), nEvents);
	
	// Export the generated events to the host memory
	phsp.Export(MyEvents);
	
	// J/psi daugthers {mu+, mu-} 
	vector<GReal_t> massesJpsi={0.1056583745, 0.1056583745};
	
	// Constructor
	PhaseSpace phspJpsi( 3.096916, massesJpsi, nEvents);
	
	// Pass the J/psi's from the B0 decay.
	phspJpsi.Generate( phspB0.GetDaughters(0)  );
	
	Events *MyEventsJpsi = new Events(massesJpsi.size(), nEvents );
	phspJpsi.Export(MyEventsJpsi);
\end{lstlisting}
\end{center}

Typical C++ code to generate and export to the 
host memory the 
$B^0\to K^+\pi^- {J/\psi}$ {events},
with ${J/\psi}\to\mu^+\mu^-$ is shown 
in \autoref{lst:BKPiJpsi_snipet}.
The comments in the listing explain the details about each line. 
Basically, the class PhaseSpace is instantiated and configured to 
generate the first level of the decay chain $B^0\to K^+\pi^- 
{J/\psi}$. 
The  $B^0$s are at rest.
The $B^0\to K^+\pi^- {J/\psi}$ events are then 
copied to the host memory. 
In the next step, a second PhaseSpace object is instantiated
and configured to generate the ${J/\psi}\to\mu^+\mu^-$ decay. 
The generation is performed invoking the \cobj{PhaseSpace::Generate} 
method and passing as argument 
the list of ${J/\psi}$ daughters taken from the 
$B^0\to K^+\pi^- {J/\psi}$ decay. 
The generated ${J/\psi}\to\mu^+\mu^-$ 
decays are copied to the host 
memory as well.

\subsection{Evaluation of function over the events }

\begin{center}
\begin{lstlisting}[caption={C++ code listing showing the stateless functor Dataset, which is implements the calculation of the five variables described \autoref{sec:worked_example}.} ,
 label=lst:functor_snipet, float,basicstyle=\tiny,linewidth=.98\textwidth]
struct Dataset: public IFunctionArray
{
	__host__ __device__
	GReal_t cosHELANG(Vector4R p, Vector4R q, Vector4R d )
	{ // details of the calculation omited for brevity }

	__host__ __device__
	GReal_t deltaAngle(const Vector4R& p4_p,
			const Vector4R& p4_d1,	const Vector4R& p4_d2,
			const Vector4R& p4_h1,	const Vector4R& p4_h2 )
	{// details of the calculation omited for brevity }

	__host__ __device__
	void  operator()(const GInt_t n, Vector4R** particles,
	GReal_t* variables  )
	{
		Vector4R pJpsi = *particles[0]; 
		Vector4R pK    = *particles[1];
		Vector4R ppi   = *particles[2];	
		Vector4R pMup  = *particles[3];
		Vector4R pMum  = *particles[4];

		Vector4R pB0   =  pJpsi + pK +ppi; Vector4R pKpi  =  pK    + ppi;
		
		variables[0] = pKpi.mass(); 
		variables[1] = cosHELANG( pB0, pKpi, pK );
		variables[2] = cosHELANG( pB0, pJpsi, pMup ) ;
		variables[3] = deltaAngle( pB0, pK, ppi , pMup, pMum ) ;
	}

};
\end{lstlisting}
\end{center}

Using \cobj{MCBooster::Evaluate} and \cobj{MCBooster::EvaluateArray} routines,
it is possible to evaluate arbitrary function objects in parallel, taking events stored in
the device memory as parameters. The results can be kept in the device or copied automatically to the host memory 
depending on the parameters passed to the algorithms.
{\autoref{lst:dataset_snipet}} {shows}
 how to make a dataset with four parameters, 
as discussed in  \autoref{sec:worked}.
The results of the calculation are stored in the host memory.

\begin{center}
\begin{lstlisting}[caption={C++ code listing showing how to evaluate a functor and build up a dataset with five variables running in parallel
over 10 million events of $B^0 \to K^+\pi^- J\psi$ stored in device memory.} ,
 label=lst:dataset_snipet, float, basicstyle=\tiny,linewidth=.98\textwidth]
        // Set of variables. Each element points to an array
        // of doubles in device memory
	VariableSet_h Var(4);
	RealVector_h result_MKpi(events);
	RealVector_h result_CosThetaK(events);
	RealVector_h result_CosThetaMu(events);
	RealVector_h result_DeltaAngle(events);

	Var[0]  = &result_MKpi;
	Var[1]  = &result_CosThetaK;
	Var[2]  = &result_CosThetaMu;
	Var[3]  = &result_DeltaAngle;
        
        // Set of particles that will be taken 
        // as parameters in the object evaluation.
        // Each element points to the array of final
        // states particles that was generated previously
        // and is stored in device memory
	ParticlesSet_d JpsiKpiMuMu(5);
	JpsiKpiMuMu[0]   = &phsp.GetDaughters(0);
	JpsiKpiMuMu[1]   = &phsp.GetDaughters(1);
	JpsiKpiMuMu[2]   = &phsp.GetDaughters(2);
	JpsiKpiMuMu[3]   = &phspJpsi.GetDaughters(0);
	JpsiKpiMuMu[4]   = &phspJpsi.GetDaughters(1);

	Dataset DataJpsiKpi = Dataset();
	EvaluateArray<Dataset>(DataJpsiKpi, JpsiKpiMuMu, Var );
\end{lstlisting}
\end{center}

\section{Performance}
\label{sec:perf}

\begin{table}[t]
\begin{tabular}{c|cccc}
                       &  \multicolumn{4}{c}{Time spent per task for different cards (s)}\\            
                       &  K2200 & GTX 970 & GTX TITAN Z &  K40c  \\
\midrule
 $B^0\to {K^-} \pi^+ {J/\psi}$ & 0.034               
   & 0.016 & 0.020 & 0.021 \\
 ${J/\psi}\to\mu^+\mu^-$  & ${12\times10^{-06}}$ & 
  ${{8.8}\times10^{-06}}$ & 
  ${{9.5}\times10^{-06}}$ & 
  ${14\times10^{-06}}$ \\
 Dataset               & 0.18                & 0.10 & 0.078  & 0.049 \\
\bottomrule
\end{tabular}
\caption{The time spent, in seconds, by different NVidia GPU models to 
process 10 million events.
The row labeled $ B^0 \to K^- \pi^+ J/\psi $ is the time taken on the device
to generate this three-body decay.
The row labeled $ J / \psi \to \mu^+ \mu^- $ is the time taken on the device to
generate this two-body decay.
The row labeled Dataset is the time taken to do the calculations illustrated in
\autoref{lst:BKPiJpsi_snipet}.
} 
\label{tab:cuda}
\end{table}

{To} evaluate the performance of MCBooster, 
the time spent to perform the {various}
 operations has been measured
running on different NVidia GPUs and on a multicore CPU deploying 
different number of 
OpenMP threads. 

The performance measurements {made with} four 
CUDA{-}enabled devices with 
different {architectures} are summarized in \autoref{tab:cuda}.
Different parameters concerning each NVidia {graphics} card 
{need} to be considered to understand these numbers. 
The main parameters are listed below 

\begin{itemize}
 \item Quadro K2200 (Maxwell/5.0): 640 CUDA Cores @ 1.12 GHz.  FP64/FP32  = 1:32 
 \item GeForce GTX 970 (Maxwell/5.2): 1664 CUDA Cores @ 1.18 GHz.  FP64/FP32  = 1:32 
 \item GeForce Titan Z (Kepler/3.5): 2880 CUDA Cores @ 0.88 GHz.  FP64/FP32  = 1:3
 \item Tesla K40c (Kepler/3.5): 2880 CUDA Cores @ 0.75 GHz.  FP64/FP32  = 1:3
\end{itemize}

\begin{figure}[t]
 \begin{center}
 \includegraphics[width=.5\linewidth]{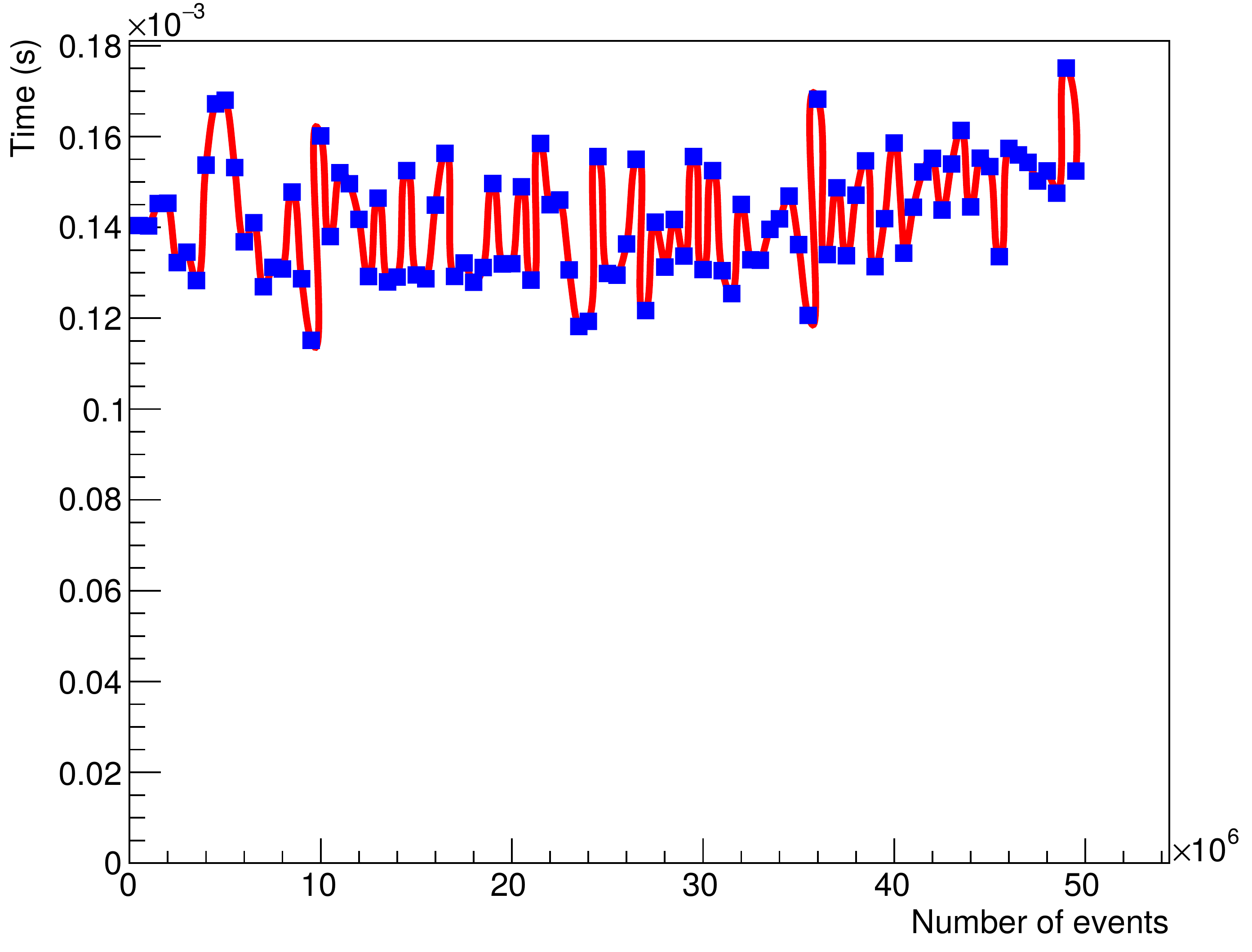}
 \end{center}
 \caption{ 
{The time spent, in seconds,
 to generate  1 million $B^0\to K^- \pi^+ J/\psi$ 
 decays on a Tesla K40c device,
 as a  function of the total number of events generated.}
} 
  \label{fig:plot1}
\end{figure}

\begin{figure}[t]
 \begin{center}
 \includegraphics[width=.5\linewidth]{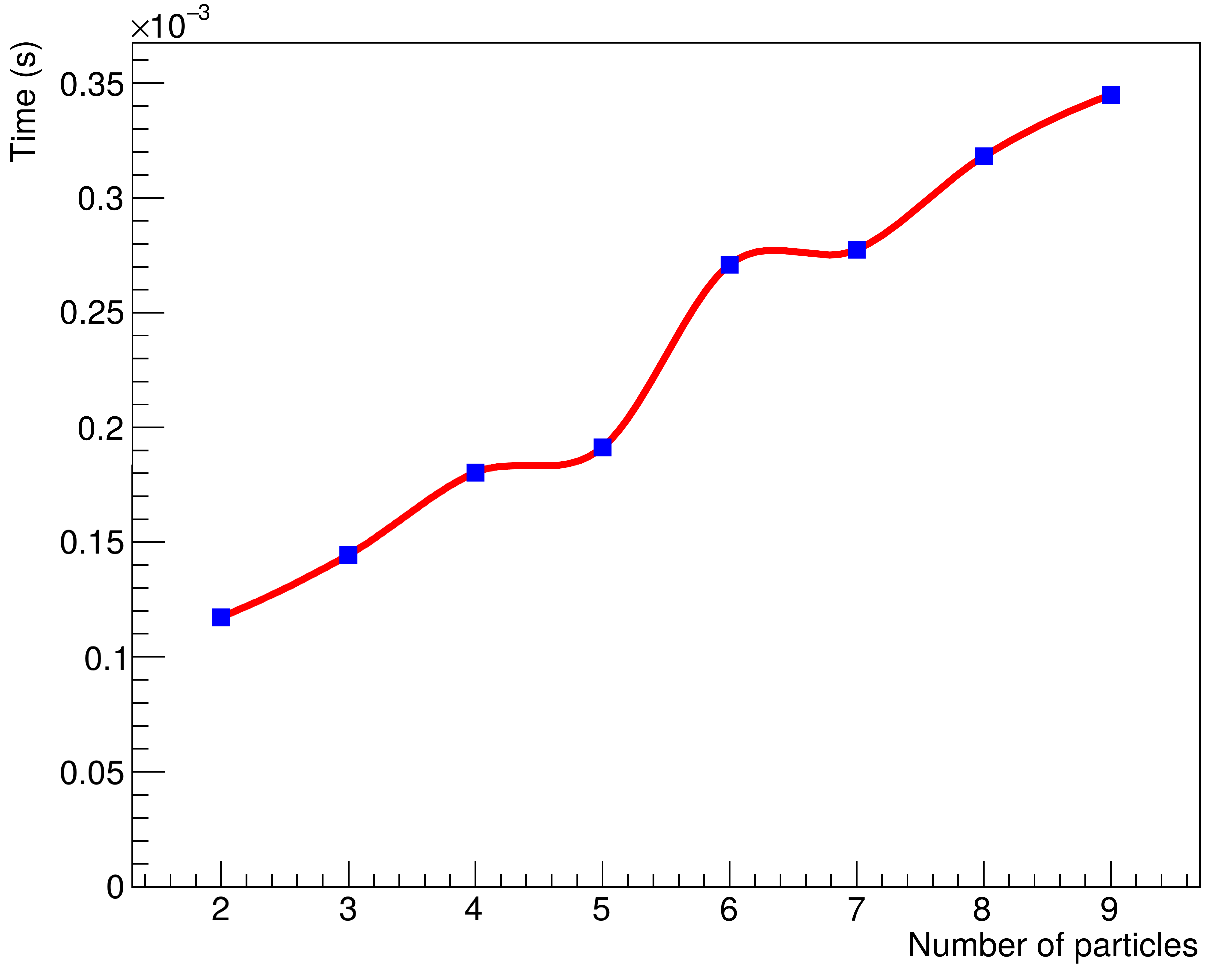}
 \end{center}
 \caption{ The time spent, in seconds, on a Tesla K40c device to 
 generate {1 million phase-space} decays 
 {as a} function {of} the number of 
  particles in the final state.}
 \label{fig:plot2}
\end{figure}

The performance variation {as a} function of  problem size was 
measured using the Tesla K40c device. 
Figure \ref{fig:plot1} shows that the time taken to generate the
underlying three-body phase-space decays is essentially independent of the number
of events between 500 thousand and 50 million events, and then begins to rise very gently.
The (small absolute) overhead setting up the problem completely dominates 
at lower statistics, and continues to dominate, even at the highest statistics
considered.
Figure \ref{fig:plot2} shows that the time to generate a sample grows roughly
linearly as a function of the number of particles in the final state.
This is the expected behavior as the Raubold-Lynch algorithm adds one recursive 
step for each additional final state particle.

The performance using 
{the} OpenMP back-end was measured using a 
{ 24 physical-core, 48 logical-core}  Xeon ES-2680 @2.5 GHz CPU. 
The results are summarized
in the \autoref{tab:openmp}.
As a further figure of {merit to describe} the performance 
{using the} OpenMP {back-end}, 
the time spent to generate
a 10 million event sample with 9 particles in final state was also measured:
\begin{itemize}
 \item MCBooster takes 0.74 seconds using 24 OpemMP threads. 
 \item TGenPhaseSpace takes 22 seconds. 
\end{itemize}
{One sees that MCBooster provides an order of magnitude performance
boost on such a system, even though it was optimized for execution on
NVidia GPUs.}

\begin{table}[t]
\begin{tabular}{c|cccc}
                       &      \multicolumn{4}{c}{Time spent per number of threads (s)}\\            
                       &  \#1   & \#12  & \#24 &  \#48  \\
\midrule
 $B^0\to K\pi^+ J/\psi$ & 3.65 & 0.369  & 0.218 & 0.161 \\
 $J/\psi\to\mu^+\mu^-$  & 2.39 & 0.232  & 0.152 & 0.136 \\
 Dataset               & 1.69 & 0.33   & 0.33  & 0.271 \\
\midrule
$B^0\to K\pi^+ J/\psi$ w/ TGenPhaseSpace   & 4.68 & -   &   -  & - \\
\bottomrule
\end{tabular}
\caption{ The time spent, in seconds,  to generate 10 million $B^0\to K^- \pi^+ J/\psi$
events under OpenMP using a 24 physical-core,  48 logical-core Xeon ES-2680 @2.5 GHz. 
\cobj{ROOT::TGenPhaseSpace} runs in a single thread.} 
\label{tab:openmp}
\end{table}

\section{Summary}

The basic design, performance and functionality of the header-only, C++11-compliant library MCBooster
have been introduced. The basic interfaces are discussed through the working example presented \label{sec:worked_example}. 
The performance measurements for running MCBooster using CUDA and OpemMP back-ends are discussed in \autoref{sec:perf}, and show that MCBooster can be 
up to 100 times faster than conventional software, depending on the graphics card or number of threads deployed. 
Since MCBooster is header only, no additional building process needs be done beyond the inclusion of the required
headers. One example of integration and usage of MCBooster can be found the on poster \cite{hasse}.

MCBooster is open source. 
The project is hosted on GitHub at  

\smallskip
\noindent
 \hspace{10mm} \url{https://github.com/MultithreadCorner/MCBooster}.

\smallskip\noindent
The package includes a suite of examples and also a simple application to generate 
samples and save {them} in ROOT TTrees. 

\section{Acknowledgments}

This work was performed with support from NSF Award PHY-1414736. 
NVidia  provided K40 GPUs for our use through its University Partnership program.
We are pleased to acknowledge valuable technical support from Christoph Hasse
and Brad Hittle.

\bibliography{references}

\end{document}